\documentclass[12pt]{article}
\usepackage{cite}
\usepackage{CJK}
\usepackage{amssymb}
\usepackage{amsmath}
\usepackage{enumerate}
\usepackage{graphicx}
\usepackage{amsfonts}
\usepackage{url}
\usepackage{bm}

\usepackage{pifont}
\usepackage{epsfig,subfigure,dsfont,amsthm,amsbsy,mathrsfs,amscd}

\def\la{\langle}
\def\ra{\rangle}
\def\be{\begin{equation}}
\def\ee{\end{equation}}
\def\ba{\begin{array}}
\def\ea{\end{array}}
\input amssym.def
\newcommand\btd{\raise 2pt \hbox{$\hat\bigtriangledown$}\hskip 1.5pt}
\newcommand\bt{\raise 2pt \hbox{$\bigtriangledown$}\hskip 1.5pt}
\begin{document}
 \title{\large\bf Maximal violation of Bell inequalities under local filtering}
\author{Ming Li$^\dag$, Huihui Qin$^\ddag$$^\S$, Jing Wang$^\dag$, Shao-Ming Fei$^\S$$^\sharp$, and Chang-Pu Sun$^\flat$\\[10pt]
\footnotesize \small $^\dag$College of the Science, China University
of Petroleum,\\
\small Qingdao 266580, P. R. China\\
\footnotesize
\small $^\ddag$Department of Mathematics, School of Science, South China University of Technology\\
\small Guangzhou 510640, P. R. China\\
\small $^\S$Max-Planck-Institute for Mathematics in the Sciences,\\
\small Leipzig 04103, Germany\\
\small $^\sharp$School of Mathematical Sciences, Capital Normal University,\\
\small Beijing 100048, P. R. China\\
\small $^\flat$ Beijing Computational Science Research Center,\\
\small Beijing 100048, P. R. China\\}
\date{}

\maketitle

\centerline{$^\ast$ Correspondence to liming@upc.edu.cn}
\bigskip

\begin{abstract}

We investigate the behavior of the maximal violations of the CHSH inequality and V$\grave{e}$rtesi's inequality under the local filtering operations.
An analytical method has been presented for general two-qubit systems to compute the maximal violation of the CHSH inequality and the lower bound of the maximal violation of V$\acute{e}$rtesi's inequality over the local filtering operations. We show by examples that there exist quantum states whose non-locality can be revealed after local filtering operation by the V$\acute{e}$rtesi's inequality instead of the CHSH inequality.
\end{abstract}
\bigskip

Quantum mechanics is inherently nonlocal. After performing local measurements on a composite quantum system, non-locality,
which is incompatible with local hidden variable theory \cite{bell} can be revealed by Bell inequalities.
The non-locality is of great importance both in understanding the
conceptual foundations of quantum theory and in investigating
quantum entanglement. It is also closely related to
certain tasks in quantum information processing, such as building
quantum protocols to decrease communication complexity \cite{dcc,dcc1}
and providing secure quantum communication \cite{scc1,scc2}. We refer to \cite{brunner} for more details.

To determine whether a quantum state has non-locality, it is sufficient to construct a Bell inequality\cite{chsh,vbell1,vbell2,vbell3,chenjingling,liprl,yu} which can be violated by the quantum state.
For two qubits systems,  Clauser-Horne-Shimony-Holt have presented the famous CHSH inequality \cite{chsh}.

Let ${\mathcal{B}_{CHSH}}$ denote the Bell operator for the CHSH inequality,
\be\label{b22}
{\mathcal{B}_{CHSH}}=A_1\otimes B_1+A_1\otimes B_2+A_2\otimes B_1-A_2\otimes B_2,
\ee
with $A_i$ and $B_j$ being the observables of the form $A_i=\sum_{k=1}^3a_{ik}\sigma_k$ and $B_j=\sum_{l=1}^3b_{jl}\sigma_l$ respectively, $i,j=1,2$,
\be\sigma_1=\left(%
    \begin{array}{cc}
      -1 & 0  \\
      0 & 1  \\
       \end{array}%
    \right),~~
\sigma_2=\left(%
    \begin{array}{cc}
      0 & 1  \\
      1 & 0  \\
       \end{array}%
    \right)\mbox{ and }
    \sigma_3=\left(%
    \begin{array}{cc}
      0 & i  \\
      -i & 0  \\
       \end{array}%
    \right)
\ee
are the Pauli matrices. For any two-qubit quantum state $\rho$, the maximal violation of the CHSH inequality (MVCI)
is given by \cite{ho340}
\be\label{maxv}
\max_{\mathcal{B_{CHSH}}}|\la{\mathcal{B_{CHSH}}}\ra_{\rho}|=2\sqrt{\tau_1+\tau_2},
\ee
where $\tau_1$ and $\tau_2$ are the two largest eigenvalues of the matrix $T^{\dag}T$, $T$ is the matrix with
entries $T_{\alpha\beta}=tr[\rho\,\sigma_\alpha\otimes\sigma_\beta]$, $\alpha,\,\beta=1,2,3$. For a state admitting local
hidden variable (LHV) model, one has $\max_{\mathcal{B_{CHSH}}}|\la{\mathcal{B_{CHSH}}}\ra_{LHV}|\leq2$.

Another effective Bell inequality for two-qubit system is given by the Bell operator \cite{vertesi} V$\acute{e}$rtesi
\be\label{ver}
\mathcal{B_V}=\frac{1}{n^2}[\sum_{i,j=1}^{n}A_i\otimes B_j+\sum_{1\leq i < j\leq
n}C_{ij}\otimes(B_i-B_j)+\sum_{1\leq i < j\leq n}(A_i-A_j)\otimes
D_{ij}], \ee where $A_i, B_j, C_{ij}$ and $D_{ij}$ are observables
of the form $\sum_{\alpha=1}^3 x_{\alpha}\sigma_{\alpha}$ with
$\vec{x}=(x_{1},x_{2},x_{3})$ the unit vectors.

The maximal violation of V$\acute{e}$rtesi's inequality(MVVI) is lower bounded by the following inequality \cite{srep}.
For arbitrary two-qubit quantum state $\rho$, we have
\begin{eqnarray}\label{ot1}
\max_{\mathcal{B_{V}}}|\la{\mathcal{B_{V}}}\ra_{\rho}|&\geq&\max_{a,b,c,d}
\left[\frac{1}{s_{ab}s_{cd}}|\int_{\Omega_a^b\times\Omega_c^d}<\vec{x},T\vec{y}>d\mu(\vec{x})d\mu(\vec{y})|
+\frac{1}{2s^2_{cd}}\int_{\Omega_c^d\times\Omega_c^d}|T(\vec{x}-\vec{y})|d\mu(\vec{x})d\mu(\vec{y})\nonumber\right.\\
&&\left.+\frac{1}{2s^2_{ab}}\int_{\Omega_a^b\times\Omega_a^b}|T^{\dag}(\vec{x}-\vec{y})|d\mu(\vec{x})d\mu(\vec{y})\right],
\end{eqnarray}
where $s_{\alpha\beta}=\int_{\Omega_{\alpha}^{\beta}}d\mu(\vec{x})$. The
maximum on the right side of the inequality goes over all the
integral area $\Omega_a^b\times\Omega_c^d$ with $0\leq a < b\leq
\frac{\pi}{2}$ and $0\leq c < d\leq \frac{\pi}{2}$. Here the maximal value $\max_{\mathcal{B_{V}}}|\la{\mathcal{B_{V}}}\ra_{\rho}|$ of a state $\rho$ admitting LHV model is upper bounded by $1$.

The maximal violation of a Bell inequality above is derived by optimizing the observables for a given quantum state.
With the formulas (\ref{maxv}) and (\ref{ot1}) one can directly check if a two-qubit quantum state violates the CHSH or the V$\acute{e}$rtesi's inequality. It has been shown that the maximal violation of a Bell inequality is in a close relation with the fidelity
of the quantum teleportation \cite{pla222} and the device-independent security of quantum cryptography \cite{prl230501}.

The maximal violation of a Bell inequality can be enhanced by local filtering operations \cite{prl170401}.
In \cite{prl160402}, the authors present a class of two-qubit entangled states admitting
local hidden variable models, and show that the states after local filtering violate a Bell
inequality. Hence, there exist entangled states, the non-locality of which can be revealed
by using a sequence of measurements.

In this manuscript, we investigate the behavior of the maximal violations of the CHSH inequality and V$\acute{e}$rtesi's inequality under local filtering operations.
An analytical method has been presented for any two-qubit system to compute the maximal violation of the CHSH inequality and the lower bound of the maximal violation of V$\acute{e}$rtesi's inequality under local filtering operations. The corresponding optimal local filtering operation is derived. We show by examples that there exist quantum states whose nonlocality can be revealed after local filtering operation by V$\acute{e}$rtesi's inequality instead of the CHSH inequality.

\medskip
\noindent{\bf Results}
\medskip

We consider the CHSH inequality for two-qubit systems first.
Before the Bell test, we apply the local filtering operation on a state
$\rho\in{\cal {H}}={\cal {H}}_{A}\otimes {\cal
{H}}_{B}$ with $dim\,{\cal {H}}_{A}=dim\,{\cal {H}}_{B}=2$. $\rho$
is mapped to the following form under local filtering
transformations \cite{verstraete,prl160402}:
\begin{eqnarray}\label{rhof}
\label{FT}\rho'=\frac{1}{N}(F_{A}\otimes
F_{B})\rho(F_{A}\otimes F_{B})^{\dag},
\end{eqnarray}
where $N=tr[(F_{A}\otimes
F_{B})\rho(F_{A}\otimes F_{B})^{\dag}]$ is a normalization factor, and $F_{A/B}$ are positive operators acting on the subsystems respectively. Such operations can be a local interaction with the dichroic environments\cite{gisinpla210}.

For two-qubit systems, let $F_{A}=U\Sigma_AU^{\dag}$ and $F_{B}=V\Sigma_BV^{\dag}$ be the spectral decompositions of $F_{A}$ and $F_{B}$ respectively,
where $U$ and $V$ are unitary operators. Define that
\be
\label{deteta}\delta_k=\Sigma_A \sigma_k \Sigma_A,~~~ \eta_l=\Sigma_B \sigma_l \Sigma_B
\ee
and $X$ be a matrix with entries given by
\be\label{defxx} x_{kl}=tr[\varrho \delta_k\otimes \eta_l],~~~ k,l=1,2,3,
\ee
where $\varrho$ is locally unitary with $\rho$.

we have the following
theorem.

{\bf{Theorem 1:}} The maximal quantum bound of a two-qubit quantum state $\rho'=\frac{1}{N}(F_{A}\otimes
F_{B})\rho(F_{A}\otimes F_{B})^{\dag}$ is given by
\be\label{lfmaxv}
\max_{\mathcal{B_{CHSH}}}|\la{\mathcal{B_{CHSH}}}\ra_{\rho'}|=\max_{\varrho}2\sqrt{\tau'_1+\tau'_2},
\ee
where $\tau'_1$ and $\tau'_2$ are the two largest eigenvalues of the matrix $X^{\dag}X/N^2$ with $X$ given by (\ref{defxx}). The left max is taken over all $B_{CHSH}$ operators, while the right max is taken over all $\varrho$ that are locally unitary equivalent to $\rho$.

See Methods for the proof of theorem 1.

Now we investigate the behavior of the V$\grave{e}$rtesi-Bell inequality under local filtering operations.
In \cite{srep} we have found an effective lower bound for the MVVI by considering infinite many
measurements settings, $n\to\infty$. Then the discrete summation in
(\ref{ver}) is transformed into an integral of the spherical
coordinates over the sphere $S^2\subset{R^3}$. We denote the
spherical coordinate of $S^{2}$ by $(\phi_1,\phi_2)$. A unit vector
$\vec{x}=(x_{1},x_{2},x_{3})$ can be parameterized by
$x_1=\sin{\phi_1}\sin{\phi_2}$,
$x_2=\sin{\phi_1}\cos\phi_{2}$, $x_3=\cos{\phi_1}$. For any $0\leq
a\leq b\leq \frac{\pi}{2},$ we denote $\Omega_{a}^b=\{x\in S^{2}:
a\leq \phi_1(x)\leq b\}.$

{\bf{Theorem 2:}} For two-qubit quantum state $\rho'$ given
by (\ref{rhof}), we have
\begin{eqnarray}\label{t1}
\max_{\mathcal{B_{V}}}|\la\mathcal{B_{V}}\ra_{\rho'}|&&\geq\max_{a,b,c,d}
\frac{1}{N}\left[\frac{1}{s_{ab}s_{cd}}|\int_{\Omega_a^b\times\Omega_c^d}<\vec{x},X\vec{y}>d\mu(\vec{x})d\mu(\vec{y})|
\nonumber\right.\\
+\frac{1}{2s^2_{cd}}&&\left.\int_{\Omega_c^d\times\Omega_c^d}|X(\vec{x}-\vec{y})|d\mu(\vec{x})d\mu(\vec{y})
+\frac{1}{2s^2_{ab}}\int_{\Omega_a^b\times\Omega_a^b}|X^t(\vec{x}-\vec{y})|d\mu(\vec{x})d\mu(\vec{y})\right],
\end{eqnarray}
where $X$ is defined by (\ref{defxx}). $X^t$ stands for the transposition of $X$, and
$s_{\alpha\beta}=\int_{\Omega_{\alpha}^{\beta}}d\mu(\vec{x})$. The
maximization on the right side of the inequality goes over all the
integral area $\Omega_a^b\times\Omega_c^d$ with $0\leq a < b\leq
\frac{\pi}{2}$ and $0\leq c < d\leq \frac{\pi}{2}$.

See Methods for the proof of theorem 2.

{\bf{Remark:}} The right hand sides of (\ref{lfmaxv}) and (\ref{t1}) depend just on the state $\sigma$ which is local unitary equivalent to $\rho$.
Thus to compare the difference of the maximal violation for $\rho$ and that for $\rho'$, it is sufficient to just consider the difference between $\sigma$ and $\rho'$.

Without loss of generality, we set
\be \Sigma_A=\left(%
    \begin{array}{cc}
      x & 0  \\
      0 & 1  \\
       \end{array}%
    \right)~\mbox{ and }~~
\Sigma_B=\left(%
    \begin{array}{cc}
      y & 0  \\
      0 & 1  \\
       \end{array}%
    \right)
\ee
with $x, y\geq 0$.
According to the definition of $\delta_k$ and $\eta_l$ in (\ref{deteta}), one computes that
\be \delta_1=\left(%
    \begin{array}{cc}
      -x^2 & 0  \\
      0 & 1  \\
       \end{array}%
    \right),\ \ \ \
\delta_2=\left(%
    \begin{array}{cc}
      0 & x  \\
      x & 0  \\
       \end{array}%
    \right)\mbox{ and }
    \delta_3=\left(%
    \begin{array}{cc}
      0 & ix  \\
      -ix & 0  \\
       \end{array}%
    \right);
\ee
\be \eta_1=\left(%
    \begin{array}{cc}
      -y^2 & 0  \\
      0 & 1  \\
       \end{array}%
    \right),\ \ \ \
\eta_2=\left(%
    \begin{array}{cc}
      0 & y  \\
      y & 0  \\
       \end{array}%
    \right)\mbox{ and }
    \eta_3=\left(%
    \begin{array}{cc}
      0 & iy  \\
      -iy & 0  \\
       \end{array}%
    \right).
\ee
Let $ \sigma_0=\left(%
    \begin{array}{cc}
      1 & 0  \\
      0 & 1  \\
       \end{array}%
    \right).$
Set $\vec{\delta}=(\delta_1,\delta_2,\delta_3)$, $\vec{\eta}=(\eta_1,\eta_2,\eta_3)$, and $\vec{\sigma}=(\sigma_0, \sigma_1,\sigma_2,\sigma_3)$.
We have $\vec{\delta}=C\vec{\sigma}$ and $\vec{\eta}=D\vec{\sigma}$, where
\be C=\left(%
    \begin{array}{cccc}
      \frac{1}{2}(1-x^2) & \frac{1}{2}(1+x^2) & 0 & 0 \\
      0 & 0  & x  & 0  \\
      0 & 0  & 0  & x  \\
       \end{array}%
    \right) {\rm and}\quad D=\left(%
    \begin{array}{cccc}
      \frac{1}{2}(1-y^2) & \frac{1}{2}(1+y^2) & 0 & 0 \\
      0 & 0  & y  & 0  \\
      0 & 0  & 0  & y  \\
       \end{array}%
    \right) {\rm respectively}.
\ee

Then one has $x_{kl}=(CWD^{\dag}),$ where $W$ is a $4\times 4$ matrix with entries $w_{\alpha\beta}=tr[\sigma\sigma_{\alpha}\otimes\sigma_{\beta}]$.
Let $\tilde{O}_A=\left(%
    \begin{array}{cc}
      1 & 0  \\
      0 & O_A  \\
       \end{array}%
    \right)$ and $\tilde{O}_B=\left(%
    \begin{array}{cc}
      1 & 0  \\
      0 & O_B  \\
       \end{array}%
    \right)$ where $O_A$ and $O_B$ are $3\times 3$ orthogonal operators. Define that $\vec{r}$ and $\vec{s}$ be three dimensional vectors with entries $r_i=tr[\rho\sigma_0\otimes\sigma_i]$ and $s_j=tr[\rho\sigma_j\otimes\sigma_0]$ respectively. And let $\tilde{T}=\left(%
    \begin{array}{cc}
      1 & \vec{r}  \\
      \vec{s} & T  \\
       \end{array}%
    \right).$
One can further show that \be\label{xcwd} X=CWD^{\dag}=C\tilde{O}_A\tilde{T}\tilde{O}_B^{\dag}D^{\dag},\ee
and \be\label{nnn}N=x_+y_++4x_-y_+(O_A\vec{s})_1+4x_+y_-(O_B\vec{r})_1+4x_-y_-(O_ATO_B^t)_{11},\ee where $x_+=\frac{1}{2}(1+x^2)$, $x_-=\frac{1}{2}(1-x^2)$,
$y_+=\frac{1}{2}(1+y^2)$ and $y_-=\frac{1}{2}(1-y^2)$. Numerically, one can parameterize $O_A$ and $O_B$ and then search for the maximization in theorem 1. For the lower bound in theorem 2, we refer to \cite{srep}.

{\bf{Corollary:}}
For two-qubit Werner state\cite{werner}
$\rho_w=p|\psi^-\ra\la\psi^-|+(1-p)\frac{I}{4}$, with
$|\psi^-\ra=(|01\ra-|10\ra)/\sqrt{2}$, one computes $T=\left(%
    \begin{array}{ccc}
    -p & 0 & 0  \\
    0 & -p & 0  \\
    0 & 0 & -p  \\
       \end{array}%
    \right).$
Then by using the symmetric property of the state, (\ref{xcwd}) and (\ref{nnn}), together with theorem 1, we have
\be\label{lfrhow}
\max_{\mathcal{B_{CHSH}}}|\la{\mathcal{B_{CHSH}}}\ra_{\rho'}|=2\sqrt{\tau'_1+\tau'_2},
\ee
where $\tau'_1$ and $\tau'_2$ are the two largest eigenvalues of the matrix $X^{\dag}X/N^2$ with $X$ given by
\be\label{defx} x_{kl}=tr[\rho_w \delta_k\otimes \eta_l],~~~ k,l=1,2,3.
\ee

\medskip
\noindent{\bf Applications}
\medskip

In the following we discuss the applications of local filtering. First we show that a state which does not
violate the CHSH and the V$\acute{e}$rtesi's inequalities could violate these inequalities after local filtering. Consider the following density matrix for two-qubit systems:
\be
\varrho_1=\frac{1}{4}(I\otimes I+r\sigma_1\otimes I-p\sum_i^3\sigma_i\otimes \sigma_i),
\ee
where $-0.3104\leq p\leq0.7$ to ensure the positivity of $\varrho_1$.
By using the positive partial transposition criteria one has that
$\varrho_1$ is separable for $-0.3104\leq p\leq0.3104$.

Case 1: Set $r=0.3$. It is direct to verify that both the CHSH inequality and V$\acute{e}$rtesi's inequalities fail to detect the non-locality for the whole region $-0.3104\leq p\leq0.7$. After filtering, non-locality can be detected for $0.6291\leq p\leq0.7$ (by Theorem 2) and $0.6164\leq p\leq0.7$ (by Theorem 1) respectively, see Fig.1.

\begin{figure}[h]
\begin{center}
\resizebox{15cm}{!}{\includegraphics{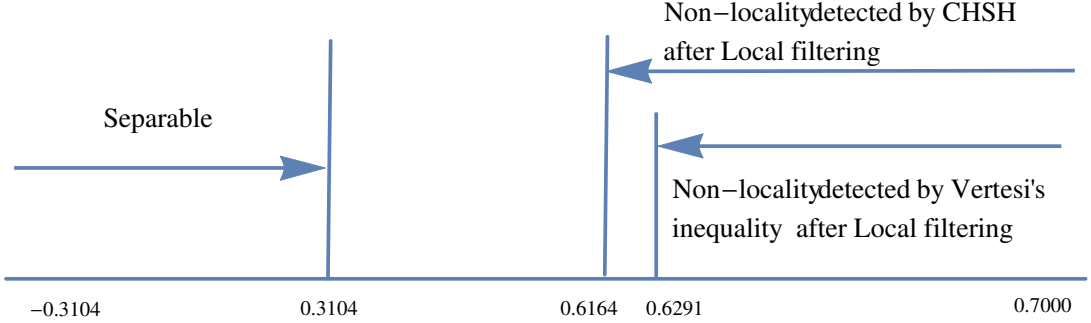}}
\end{center}
\caption{For $r=0.3$, both the CHSH inequality and V{\'e}rtesi's inequality fail to detect the non-locality of $\varrho_1$ for the whole parameter region of $p$. After local filtering, non-locality is detected for $0.6291\leq p\leq0.7$ (by Theorem 2) and $0.6164\leq p\leq0.7$ (by Theorem 1) respectively. \label{fig2}}
\end{figure}

Case 2: Set $p=0.7050$ and $r=0.0400$. The MVCI of $\varrho_1$ is $1.994$ without local filtering and $1.9988$ after local filtering, which means that the CHSH inequality is always satisfied before and after local filtering. The lower bound  (\ref{ot1}) for $\varrho_1$ is computed to be less than one, implying the non-locality can not be detected by the lower bound for MVVI derived in \cite{srep} without local filtering. However, by taking $x=y=1.1, a=c=0.1671, b=d=1.1096$, from Theorem 2 we have the maximal violation value $1.0005$ which is larger than one. Therefore, after local filtering the state's non-locality is detected.

Next we give an example that a state admits local hidden variable model (LHV) can violate the Bell inequality under local filtering. Consider two-qubit quantum states with density matrices of the following form:
\be
\varrho_2=\frac{1}{4}(I\otimes I+p\sigma_1\otimes I+p\sum_i^3\sigma_i\otimes \sigma_i).
\ee
According to the positivity of a density matrix, we have $-0.5\leq p\leq 0.3090$. By using the positive partial transposition criteria \cite{ppt}, one checks that $\varrho_2$ is entangled for $-0.5\leq p\leq -0.3090$. The quantum state satisfies the CHSH inequality for the whole parameter region.

We first show that the state $\varrho_2$ admits LHV models for $-0.5\leq p\leq -0.3090$.

First we rewrite $\varrho_2$ as a convex combination of singlet and separable states,
  \be
   \varrho_2=q|\psi_{-}\ra\la\psi_{-}|+(1-q)[\frac{1}{2}(I-\frac{q}{1-q}\sigma_{1})\otimes \frac{I}{2}],
 \ee
 where $|\psi_{-}\ra\la\psi_{-}|=\frac{1}{4}(I\otimes I-\sum^{3}_{i=1}\sigma_{i}\otimes \sigma_{i})$ and $q=-p$.
 According to \cite{Julien}, with a visibility of $q=\frac{1}{2}$, the correlations of measurement outcomes produced by measuring the observables $A=\overrightarrow{a}\cdot\overrightarrow{\sigma}$ and $B=\overrightarrow{b}\cdot\overrightarrow{\sigma}$
 on the singlet state can be simulated by an LHV model in which the hidden variable
 $\overrightarrow{\lambda}_{s} \in \mathbf{S}^{2}$ is biased distributed with probability density
 \be
 \rho(\overrightarrow{\lambda}_{s}|\overrightarrow{a})=\frac{|\overrightarrow{a}
 \cdot\overrightarrow{\lambda}_{s}|}{2\pi}.
 \ee

 With probability $0< q\leq \frac{1}{2}$, Alice and Bob can share the biased distributed variable resource
 and output $a=-sgn(\overrightarrow{a}\cdot\overrightarrow{\lambda}_{s})$ and
 $b=sgn(\overrightarrow{b}\cdot\overrightarrow{\lambda}_{s})$, respectively. With probability $1-q$, Alice outputs $a=\pm 1$ with probability $p(a|\overrightarrow{a})=tr[\frac{1}{2}(I-\frac{q}{1-q}\sigma_{z})\frac{I\pm \overrightarrow{a}
 \cdot\overrightarrow{\lambda}_{s}}{2}]$, and Bob outputs $\pm 1$ with probability $p(b|\overrightarrow{b})=\frac{1}{2}$.
  Then we can simulate the correlations produced by measuring obesrvables $A$ and $B$ on $\varrho_2$,
  \be
  p(a,b|\overrightarrow{a},\overrightarrow{b},\varrho_2)=tr(\frac{I+a\overrightarrow{a}\overrightarrow{\sigma}}{2}\otimes
  \frac{I+b\overrightarrow{b}\overrightarrow{\sigma}}{2}\rho)=\frac{1-qab\overrightarrow{a}\cdot\overrightarrow{b}}{4}
  -\frac{aa_{3}q}{4},
  \ee
which can be given by the following LHV model,
  \be
  \begin{aligned}
   p(a,b|\overrightarrow{a},\overrightarrow{b},\varrho_2)=\displaystyle&q\int_{\mathbf{S}^{2}}
   p(a|\overrightarrow{a},\overrightarrow{\lambda}_{s})
   p(b|\overrightarrow{b}\cdot\overrightarrow{\lambda}_{s})\rho(\overrightarrow{\lambda}_{s})d\overrightarrow{\lambda}_{s}+
   (1-q)p(a\rvert\overrightarrow{a})p(b\rvert\overrightarrow{b})\\
   =\displaystyle&q\int_{\Omega_{a,b}}\frac{|\overrightarrow{a}\cdot\overrightarrow{\lambda}_{s}|}{2\pi}
   d\overrightarrow{\lambda}_{s}+(1-q)p(a|\overrightarrow{a})p(b|\overrightarrow{b}),
  \end{aligned}
  \ee
  where $\Omega_{a,b}=\{\overrightarrow{\lambda}_{s}|-sgn(\overrightarrow{a}\cdot\overrightarrow{\lambda}_{s})=a\}\cap
   \{\overrightarrow{\lambda}_{s}|b=sgn(\overrightarrow{b}\cdot\overrightarrow{\lambda}_{s})\}$.
Explicitly,
  $$
  p(1,1\arrowvert\overrightarrow{a},\overrightarrow{b},\overrightarrow{\lambda}_{s})=q\int_{\Omega_{1,1}}\frac{|\overrightarrow{a}
  \cdot\overrightarrow{\lambda}_{s}|}{2\pi}d\overrightarrow{\lambda}_{s}
  +\frac{1-q}{2}tr[\frac{1}{2}(I-\frac{q}{1-q}\sigma_{z})\frac{I+\overrightarrow{a}
 \cdot\overrightarrow{\lambda}_{s}}{2}],
$$
$$
  p(1,-1\arrowvert\overrightarrow{a},\overrightarrow{b},\overrightarrow{\lambda}_{s})=q\int_{\Omega_{1,-1}}\frac{|\overrightarrow{a}
  \cdot\overrightarrow{\lambda}_{s}|}{2\pi}d\overrightarrow{\lambda}_{s}
  +\frac{1-q}{2}tr[\frac{1}{2}(I-\frac{q}{1-q}\sigma_{z})\frac{I+\overrightarrow{a}
 \cdot\overrightarrow{\lambda}_{s}}{2}],
$$
$$
  p(-1,1\arrowvert\overrightarrow{a},\overrightarrow{b},\overrightarrow{\lambda}_{s})=q\int_{\Omega_{-1,1}}\frac{|\overrightarrow{a}
  \cdot\overrightarrow{\lambda}_{s}|}{2\pi}d\overrightarrow{\lambda}_{s}
  +\frac{1-q}{2}tr[\frac{1}{2}(I-\frac{q}{1-q}\sigma_{z})\frac{I-\overrightarrow{a}
 \cdot\overrightarrow{\lambda}_{s}}{2}],
$$
$$  p(-1,-1\arrowvert\overrightarrow{a},\overrightarrow{b},\overrightarrow{\lambda}_{s})=q\int_{\Omega_{-1,-1}}\frac{|\overrightarrow{a}
  \cdot\overrightarrow{\lambda}_{s}|}{2\pi}d\overrightarrow{\lambda}_{s}
  +\frac{1-q}{2}tr[\frac{1}{2}(I-\frac{q}{1-q}\sigma_{z})\frac{I-\overrightarrow{a}
 \cdot\overrightarrow{\lambda}_{s}}{2}],
$$
where $\Omega_{1,1}=\{\overrightarrow{\lambda}_{s}\arrowvert
  \overrightarrow{a}\cdot\overrightarrow{\lambda}<0\}\cap\{\overrightarrow{\lambda}_{s}\arrowvert\overrightarrow{b}\cdot
  \overrightarrow{\lambda}\geq0\}$,
  $\Omega_{1,-1}=\{\overrightarrow{\lambda}_{s}\arrowvert
  \overrightarrow{a}\cdot\overrightarrow{\lambda}<0\}\cap\{\overrightarrow{\lambda}_{s}\arrowvert\overrightarrow{b}\cdot
  \overrightarrow{\lambda}<0\}$,
  $\Omega_{-1,1}=\{\overrightarrow{\lambda}_{s}\arrowvert
  \overrightarrow{a}\cdot\overrightarrow{\lambda}\geq0\}\cap\{\overrightarrow{\lambda}_{s}\arrowvert\overrightarrow{b}\cdot
  \overrightarrow{\lambda}\geq0\}$,
  $\Omega_{-1,-1}=\{\overrightarrow{\lambda}_{s}\arrowvert
  \overrightarrow{a}\cdot\overrightarrow{\lambda}\geq0\}\cap\{\overrightarrow{\lambda}_{s}\arrowvert\overrightarrow{b}\cdot
  \overrightarrow{\lambda}<0\}$.

Therefore the state $\varrho_2$ admits LHV model for $-0.5\leq p\leq-0.309$.
However, after local filtering, non-locality (violation of the CHSH inequality) is detected for $-0.5\leq p\leq -0.4859$, see Fig.2.

\begin{figure}[h]
\begin{center}
\resizebox{10cm}{!}{\includegraphics{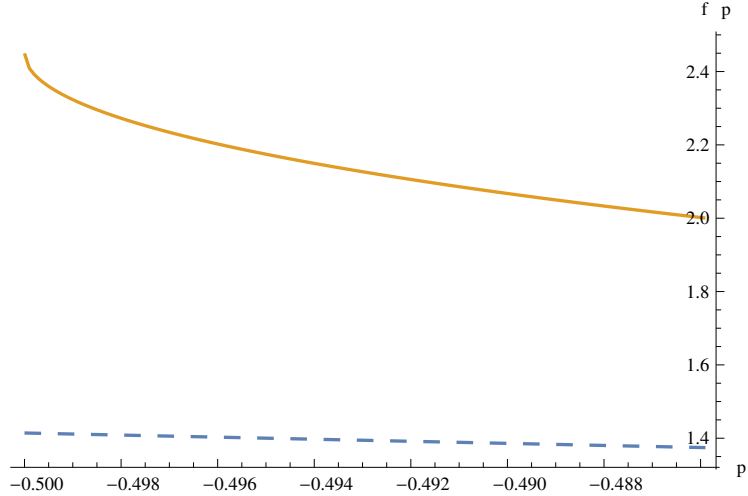}}
\end{center}
\caption{The MVCI of $\varrho_2$ (dashed line) v.s. the MVCI after Local filtering (solid line). $f(p)$ stands for the MVCI.
Note that the classical bound of the CHSH inequality is $2$.\label{fig1}}
\end{figure}

{\bf{Remark:}} In \cite{pla222} Horodeckis have presented the connection between the maximal violation of the CHSH inequality and the optimal quantum
teleportation fidelity:
\be\mathcal{F}_{max}\geq \frac{1}{2}(1+\frac{1}{12}\max_{\mathcal{B_{CHSH}}}|\la{\mathcal{B_{CHSH}}}\ra_{\rho}|)\ee
which means that any two-qubit quantum state violating the CHSH inequality is useful for teleportation and vice versa.
Ac$\acute{i}$n et al. have derived the relation between the maximal violation of the CHSH inequality and the Holevo quantity between Eve
and Bob in device-independent Quantum key distribution(QKD)\cite{prl230501}:
\be\chi(B_1:E)\leq h(\frac{1+\sqrt{(\max_{\mathcal{B_{CHSH}}}|\la{\mathcal{B_{CHSH}}}\ra_{\rho}|/2)^2-1}}{2}),\ee
where $h$ is the
binary entropy. From our theorem, $\max_{\mathcal{B_{CHSH}}}|\la{\mathcal{B_{CHSH}}}\ra_{\rho}|$ can be enhanced by implementing a proper local filtering operation from smaller to larger than $2$, which makes a teleportation possible from impossible, or can be improved to obtain a better teleportation fidelity.
The proper(optimal) local filtering operation can be selected by the optimizing process in (\ref{lfmaxv}) together with the double cover relationship between the $SU(2)$ and $SO(3)$. For application in the QKD, Eve can enhance the upper bound of Holevo quantity by local filtering operations which makes a chance for attacking the protocol.

\medskip
\noindent{\bf Discussions}

It is a fundamental problem in quantum theory to recognize and explore
the non-locality of a quantum system.
The Bell inequalities and their maximal violations supply powerful ability to detect and qualify the non-locality.
Furthermore, the constructing and the computation of the maximal violation of a Bell inequality is in close relationship with quantum games, minimal Hilbert space dimension and dimension witnesses, as well as quantum communications such as communication complexity, quantum cryptography, device-independent quantum key distribution etc. \cite{brunner}. A proper local filtering operation can generate and enhance the non-locality.
We have investigated the behavior of the maximal violations of the CHSH inequality and the V$\acute{e}$rtesi's inequality under local filtering. We have presented an analytical method for any two-qubit system to compute the maximal violation of the CHSH inequality and the lower bound of the maximal violation of V$\acute{e}$rtesi's inequality under local filtering. We have shown by examples that there exist quantum states whose nonlocality can be revealed by local filtering operations in terms of the V$\acute{e}$rtesi's inequality instead of the CHSH inequality.

\medskip
\noindent{\bf Methods}
\medskip

{\sf Proof of Theorem 1 and Theorem 2}~

The normalization factor $N$ has the following form,
\begin{eqnarray}\label{n}
N&=&tr[U\Sigma_A^2U^{\dag}\otimes V\Sigma_B^2V^{\dag}\rho]=tr[\Sigma_A^2\otimes \Sigma_B^2U^{\dag}\otimes V^{\dag}\rho U\otimes V]\nonumber\\
&=&tr[\Sigma_A^2\otimes \Sigma_B^2\varrho],
\end{eqnarray}
where $\varrho=U^{\dag}\otimes V^{\dag}\rho U\otimes V$.
Since $\rho$ and $\varrho$ are local unitary equivalent, they must have the same value of the maximal violation for CHSH inequality.

We have that
\begin{eqnarray}\label{8}
t'_{ij}&=&tr[\rho'\sigma_i\otimes\sigma_j]=\frac{1}{N}tr[(F_{A}\otimes
F_{B})\rho(F_{A}^{\dag}\otimes F_{B})^{\dag}\sigma_i\otimes\sigma_j]\nonumber\\
&=&\frac{1}{N} tr[\rho U\Sigma_A U^{\dag}\sigma_iU\Sigma_A U^{\dag}\otimes V\Sigma_B V^{\dag}\sigma_jV\Sigma_B V^{\dag}]\nonumber\\
&=&\frac{1}{N}\sum_{kl}tr[U^{\dag}\otimes V^{\dag}\rho U\otimes V \Sigma_A O^A_{ik}\sigma_k \Sigma_A\otimes \Sigma_B O^B_{jl}\sigma_l \Sigma_B]\nonumber\\
&=&\frac{1}{N}\sum_{kl}O^A_{ik}O^B_{jl}tr[\varrho \Sigma_A \sigma_k \Sigma_A\otimes \Sigma_B \sigma_l \Sigma_B]\nonumber\\
&=&\frac{1}{N}\sum_{kl}O^A_{ik}O^B_{jl}tr[\varrho \delta_k\otimes \eta_l]\nonumber\\
&=&\frac{1}{N}\sum_{kl}O^A_{ik}x_{kl}O^B_{jl}=\frac{1}{N}(O_AXO_B^T)_{ij}.
\end{eqnarray}

In deriving the fourth equality in (\ref{8}) we have used the double cover relation between the special unitary group $SU(2)$ and the special orthogonal group $SO(3)$:
for any given unitary operator $U$, $U\sigma_{i}U^{\dag}=\sum\limits_{j=1}^{3}O_{ij}\sigma_{j}$,
where the matrix $O$ with entries $O_{ij}$ belongs to $SO(3)$ \cite{4396,lilu}.

Finally, one has that
\begin{eqnarray}\label{ntt}
T'=\frac{1}{N}O_AXO_B^{\dag},
\end{eqnarray}
and
\begin{eqnarray}\label{nt}
(T')^{\dag}T'=\frac{1}{N^2}O_BX^{\dag}O_A^{\dag}O_AXO_B^{\dag}=\frac{1}{N^2}O_BX^{\dag}XO_B^{\dag}.
\end{eqnarray}

By noticing the orthogonality of the operator $O_B$ we have that the eigenvalues of $(T')^{\dag}T'$ and $X^{\dag}X/N^2$ must be the same, which proves
theorem 1.

We can further obtain theorem 2 by substituting (\ref{ntt}) into (\ref{ot1}).
\hfill \rule{1ex}{1ex}

\newpage
\bigskip
\noindent{\sf Acknowledgements}

\noindent This work is finished in the Beijing Computational Science Research Center and is supported by the NSFC Grants No. 11275131 and No. 11675113; the Shandong Provincial Natural Science Foundation No.ZR2016AQ06; the Fundamental Research Funds for the Central Universities Grants No. 15CX08011A and No. 16CX02049A; Qingdao applied basic research program No. 15-9-1-103-jch, and a project sponsored by SRF for ROCS, SEM.

\bigskip
\noindent{\sf Author contributions}

\noindent  M. Li and H.H. Qin wrote the main manuscript text. All
authors reviewed the manuscript.

\bigskip
\noindent{\sf Additional Information}

\noindent Competing Financial Interests: The authors declare no competing financial interests.

\end{document}